# Long-range Coulomb-like mode interaction in a coaxial waveguide filled with Faraday material.

**Igor I. Smolyaninov**

Department of Electrical and Computer Engineering, University of Maryland, College Park,

MD 20742

telephone:301-405-3255, fax: 301-314-9281, e-mail: smoly@eng.umd.edu

Nonlinear mode coupling in a coaxial waveguide filled with Faraday material has been considered. The picture of mode interaction is shown to resemble Coulomb interaction of charges: higher modes with nonzero angular momentum interact like effective charges via exchange of zero angular momentum quanta of the fundamental mode. Thus, at large distances this interaction becomes the dominant mechanism of mode coupling. The developed model may be used in designing coaxial photonic crystal fibers with strong tailored mode interaction.

Optical fiber communication, all-optical signal processing, and many other fields of optics require understanding of mode interaction in a guiding optical fiber or waveguide. As a result, considerable efforts have been put into the mode coupling modeling and calculations. In general, these model calculations require extensive numerical efforts and are very difficult to analyze analytically.

Very recently [1] consideration of cylindrical surface plasmon mode coupling in metal nanowires and nanoholes has led to the introduction of analytical picture of mode coupling in these systems, which strongly resembles Coulomb interaction of electric charges. In this picture higher ($n > 0$) cylindrical surface plasmon (CSP) modes are shown to posses a quantized effective charge proportional to their angular momentum $n$. In a metal nanowire these relatively slow moving effective charges exhibit strong long-range interaction via exchange of fast CSPs with zero angular momentum ($n=0$). Such a mode-coupling picture leads to a quasi one-dimensional Coulomb interaction of higher CSP modes, which becomes rather strong over long distances along the metal nanowire.

In this paper I am going to show that the geometry of electric and magnetic fields of the optical modes in a coaxial waveguide filled with Faraday material is very similar to the geometry of CSP modes in a metal nanowire. Thus, similar "effective charges" picture of nonlinear mode interaction may be built in the case of mode coupling in a coaxial waveguide filled with Faraday material. Moreover, the developed model may be used in designing coaxial photonic crystal fibers with similar picture of mode interaction. Thus, special fibers may be designed which exhibit tailored, strong, and analytically clear picture of mode interaction.

Let us first consider a coaxial waveguide (Fig.1) filled with a transparent isotropic material, which exhibits magnetic field induced optical activity (a Faraday material). In such a chiral material the relationship between the **D** and **E** vectors looks like [2]

$$\vec{D} = \varepsilon\vec{E} + i\vec{E} \times \vec{g} \qquad (1)$$

where **g** is the gyration vector proportional to the magnetic field:

$$\vec{g} = f\vec{H} \qquad (2)$$

where the constant *f* may be either positive or negative. Similar to the *n=0* cylindrical surface plasmon mode of a metal nanowire, the quanta of the fundamental mode of the coaxial waveguide may be considered as the quanta of the effective angular rotation field **Ω**:

$$\vec{\Omega} = \frac{c}{2\varepsilon}\vec{\nabla} \times \vec{g} \qquad (3)$$

(where *c* is the speed of light), which may be defined locally for any chiral medium regardless of the nature of its optical activity (natural or magnetic field induced) [1]. This statement is evident from the fact that the oscillating axially symmetric magnetic field $H_\phi$ of the fundamental mode of the coaxial waveguide corresponds to small oscillations of **Ω** (here it is also important to emphasize that the fundamental mode of the coaxial waveguide has no cut-off frequency). Thus, when a relatively slow moving compact wave packet of a higher guided mode of the coaxial waveguide (with nonzero total angular momentum **L**) experiences the action of the low frequency fundamental mode (Fig.2), such a wave packet acquires additional energy *ΔE=L**Ω***. This statement can be confirmed by considering the behavior of the solutions of Maxwell equations in the presence of the axial gyration field $g_\phi = g_\phi(r,z,t)$. After simple calculations the wave equation can be written in the form

$$-\Delta \vec{B} = -\frac{\varepsilon}{c^2}\frac{\partial^2 \vec{B}}{\partial t^2} + \frac{i}{c}\frac{\partial(\vec{\nabla}\times[\vec{E}\times\vec{g}])}{\partial t} \qquad (4)$$

which for the z-components of a solution $\sim e^{in\phi}$ can be written in the form

$$\frac{\partial}{r\partial r}\left(r\frac{\partial B_z}{\partial r}\right) + \frac{\partial^2 B_z}{\partial z^2} - \frac{\varepsilon \partial^2 B_z}{c^2 \partial t^2} - \frac{n^2}{r^2}B_z + \frac{ing}{rc}\frac{\partial(iE_z)}{\partial t} + \frac{in}{cr}\left(\frac{\partial g}{\partial t}\right)(iE_z) = 0 \qquad (5)$$

which is similar to the Klein-Gordon equation, where $n$ and $g/r$ play the role of the effective "electric" charge, and the effective potential, respectively [1] (here it is important to mention that for such higher guided modes $iE_z \sim B_z$, where the coefficient of proportionality is determined by the boundary conditions).

On the other hand, solution of the nonlinear Maxwell equations indicates that the modes with higher angular momentum (the effective charges) behave as the sources of the field of the fundamental mode. Let us search for the solutions of the nonlinear Maxwell equation (4) of the form $\boldsymbol{B}=\boldsymbol{B_0}+\boldsymbol{B_n}$ and $\boldsymbol{E}=\boldsymbol{E_0}+\boldsymbol{E_n}$, where $\boldsymbol{B_0}$ and $\boldsymbol{B_n}$ are the fundamental mode and the $n>0$ guided mode, respectively, and the gyration field is obtained in a self-consistent manner as $\boldsymbol{g}=f(\boldsymbol{B_0}+\boldsymbol{B_n})$. We are interested in the solution for the field $\boldsymbol{B_0}$ in the limit of small frequencies $\omega_0$ in the presence of $\boldsymbol{B_n}$ field, so that in the found solution the $\boldsymbol{B_n}$ field will act as a source of $\boldsymbol{B_0}$. After neglecting the terms proportional to $f^2$, and taking into account that $\boldsymbol{B_0}$ and $\boldsymbol{B_n}$ are incoherent solutions of linear Maxwell equations we obtain

$$\Delta \vec{B}_0 = \frac{4\pi\omega_n f}{c^2}\vec{\nabla}\times\vec{S}_n \qquad (6)$$

where $\boldsymbol{S_n}$ is the Pointing vector of the $n$-th mode. This equation is somewhat similar to the Poisson equation for electric charges. Moreover, using vector calculus theorem we can also derive an analog of the Gauss theorem for our effective "chiral" charges. Let us consider a cylindrical volume $V$ around the coaxial waveguide (see Fig.3), such that the side wall of the volume $V$ is located very far from the waveguide and the electromagnetic field is zero at this wall. If $S$ is the closed two-dimensional cylindrical surface bounding $V$, with area element $da$ and unit outward normal $n$ at $da$, and $S_1$ and $S_2$ are the front and the back surfaces of $V$, we can write the following integral equation for the Pointing vector $S_n$ of the $n$-th mode:

$$\int_V \vec{\nabla} \times \vec{S}_n d^3x = \int_{S2} \vec{N} \times \vec{S}_n da - \int_{S1} \vec{N} \times \vec{S}_n da \qquad (7)$$

where $N$ is the chosen direction of the coaxial waveguide. Using equation (6) we obtain

$$\int_V \frac{4\pi f \omega_n}{c^2} \vec{\nabla} \times \vec{S}_n d^3x = \int_{S2} \vec{N} \times [\vec{\nabla} \times \vec{B}_0] da - \int_{S1} \vec{N} \times [\vec{\nabla} \times \vec{B}_0] da \qquad (8)$$

Since $\vec{N} \times [\vec{\nabla} \times \vec{B}_0] = \frac{\partial B_{0\phi}}{\partial z}$, we see that a "chiral charge" produces a local step in the gyration field. Thus, the "chiral charges" interact according to the one-dimensional Coulomb law with the interaction energy growing linearly with distance. In reality this idealized linear growth is cut off by the absorption in the waveguide. However, at small frequencies of the fundamental mode the absorption is small and the range of chiral interaction is large. In fact, this unusual long-range nonlinear optical mode interaction should become the most important at large distances between the wave packets of higher modes. The numerical estimate made for a cylindrical surface plasmon case [1] indicates that large nonlinear effects may be observed.

It is also important to mention that the described mechanism of mode interaction may be realized in coaxial photonic crystal fibers in which the light is guided in the coaxial gap between the central and the outer photonic crystal structures. Introduction of metal wire inside the central photonic crystal region and the coaxial metal cylinder outside the outer photonic crystal region (Fig.4) creates the cut-off-free fundamental mode similar to the fundamental mode of a usual coaxial waveguide. On the other hand, the inner and outer photonic crystal region should limit the optical absorption of the guided modes at higher frequencies. Such a combined coaxial photonic crystal waveguide may be an extremely valuable tool in all-optical signal processing.

This work was supported in part by the NSF NER Grant No. ECS-0210438.

**Figure Captions**

Figure 1. Electric (solid lines) and magnetic (dashed lines) field distribution of the fundamental guided mode in a coaxial waveguide.

Figure 2. Interaction between the wave packet of a higher guided mode (solid line) and the low-frequency fundamental mode (dashed line) in a coaxial waveguide.

Figure 3. Schematic view of the coaxial waveguide and the auxiliary surfaces used in the derivation of the effective Gauss theorem.

Figure 4. Schematic view of the suggested coaxial photonic crystal waveguide.

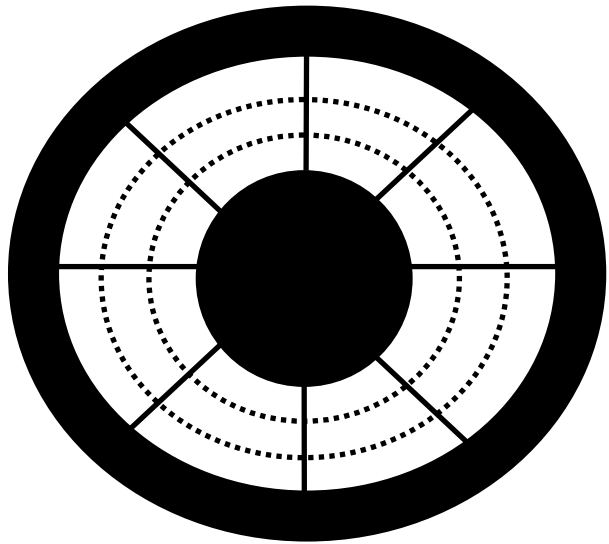

Figure 1.

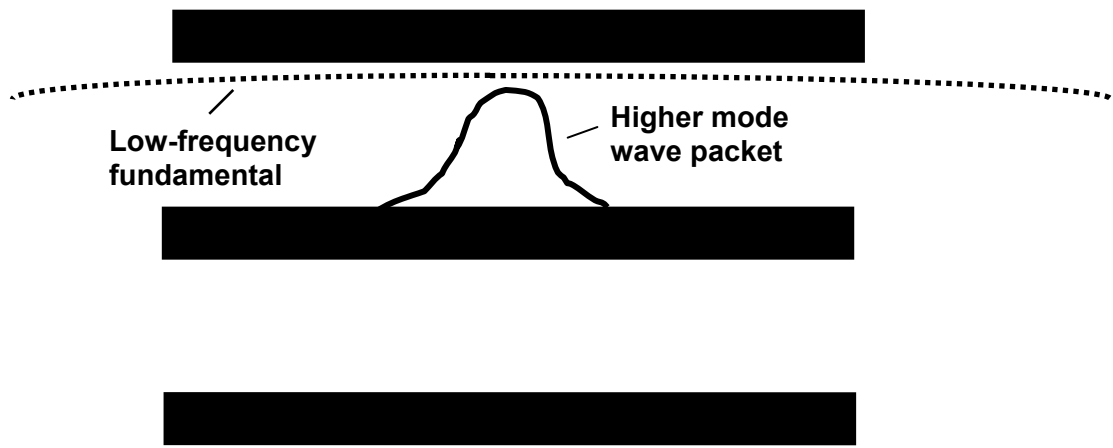

Figure 2.

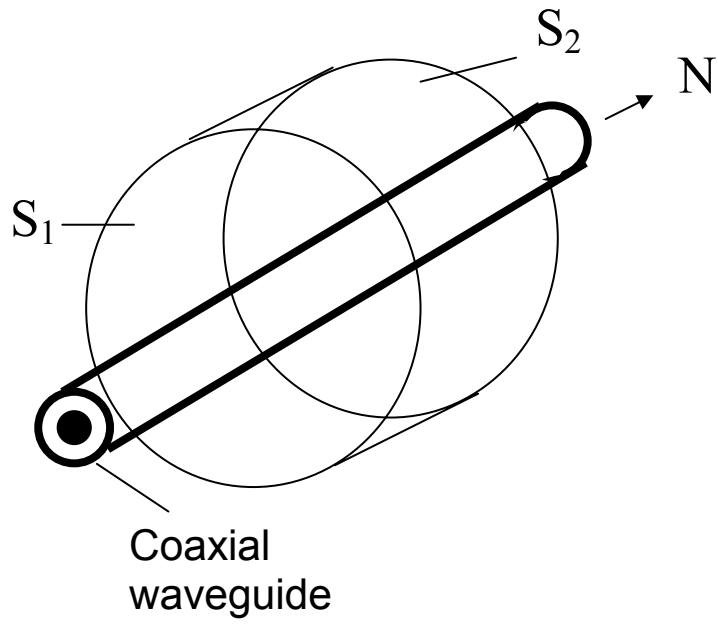

Figure 3.

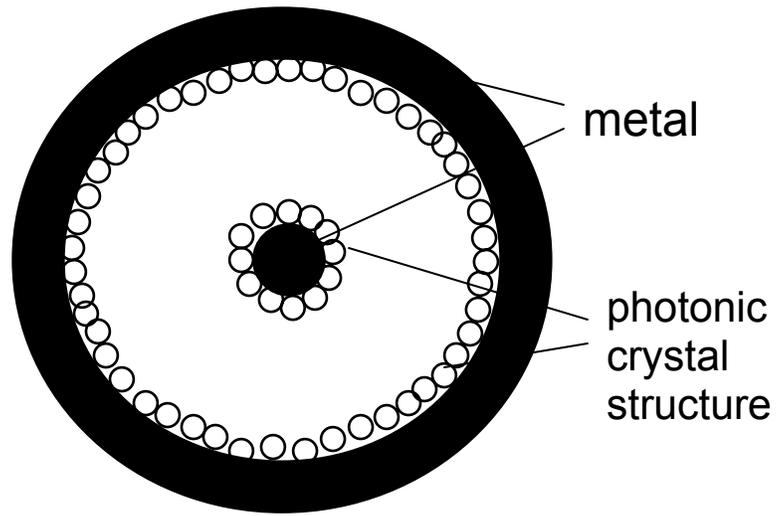

Figure 4.